# The Interconnection Tensor Rank and the Neural Network Storage Capacity

**B.V. Kryzhanovsky**
*Scientific Research Institute for System Analysis of the NRC «Kurchatov Institute», Moscow*
*ORCID: ID 0000-0002-0901-6370 , email: kryzhanov@mail.ru*

**Abstract** - Neural network properties are considered in the case of the interconnection tensor rank being higher than two (i.e. when in addition to the synaptic connection matrix, there are presynaptic synapses, pre-presynaptic synapses, etc.). This sort of interconnection tensor occurs in realization of crossbar-based neural networks. It is intrinsic for a crossbar design to suffer from parasitic currents: when a signal travels along a connection to a certain neuron, a part of it always passes to other neurons' connections through memory cells (synapses). As a result, a signal at the neuron input holds noise – other weak signals going to all other neurons. It means that the conductivity of an analog crossbar cell varies proportionally to the noise signal, and the cell output signal becomes nonlinear. It is shown that the interconnection tensor of a certain form makes the neural network much more efficient: the storage capacity and basin of attraction of the network increase considerably. A network like the Hopfield one is used in the study.

## INTRODUCTION

The formation of inter-neuron connections (synapses) is the most difficult problem in realization of an artificial neural network. Most neural network algorithms require no less than $10^5$ such interconnections, which is almost impossible to be realized electronically.

One of possible solutions to this problem is to arrange interconnections in a crossbar form [1-7]. Practicable crossbar circuits engage analog memory cells. These nonvolatile cells usually use ferroelectric [8, 9], magnetic [10], organic [11, 12] and metal-oxide [13-18] materials, floating-gate transistors [19-21], phase-change memory [22-24]. Such crossbar circuits are components of hybrid signal processing systems used for vector matrix multiplication, a basic operation of any artificial neural network. The important thing about these circuits is that they make it possible to execute physical-layer computations using fundamental laws of Ohm and Kirchhoff. Unlike digital methods, the approach allows much higher energy and space efficiency.

An essential drawback of crossbar circuits is the presence of leakage and parasitic currents. While leakages can be handled in one way or another, parasitic currents are an inherent feature of crossbar architecture: a signal going along a connection to a certain (the *j*-th) neuron spreads over all other connections via memory cells. Thus, the input signal of the *i*-th neuron holds noise – weak signals going to all other neurons. At the same time the conductivity of an analog cell changes proportionately to the noise amplitude and the cell output acquires a nonlinear relation with the input. Below expression (1) somewhat describes the situation. The presence of noise usually impairs the efficiency of a neural network. Below we show that the impairment is not unavoidable: under certain conditions this sort of noise can increase the neural network efficiency noticeably. A network of Hopfield type [25] is used in the research. This sort of neural network is thoroughly investigated [26-39], and the result of its realization can be easily compared with available analytical data.

## DESCRIPTION OF THE MODEL

Let us look at a Hopfield-type network consisting of $N$ neurons. In the network a local field acting on the $i$-th neuron (the input signal of the $i$-th neuron) is set by the expression:

$$h_i = \sum_{r=1}^{\infty} \sum_{j_1,...,j_r} T^{(r)}_{i,j_1,...,j_r} y_{j_1} y_{j_2} ... y_{j_r} \qquad (1)$$

where $y_j$ is the output signal of the $j$-th neuron ($i, j = 1, 2, ..., N$), $T^{(r)}_{i,j_1,...,j_r}$ is the neuron interconnection tensor defined below.

We assume that the associative memory is built on $M$ random binary patterns $\mathbf{x}_\mu = (x_1^{(\mu)}, x_2^{(\mu)}, ..., x_N^{(\mu)})$, $x_i^{(\mu)} = \pm 1$, $\mu = 1, 2, ..., M$. Connections of the $i$-th neuron are set in the fashion similar to the Hebb rule [40]:

$$T^{(r)}_{i,j_1,...,j_r} = a_r \sum_{\mu=1}^{M} x_i^{(\mu)} x_{j_1}^{(\mu)} ... x_{j_r}^{(\mu)} \qquad (2)$$

The difference is in that the Hebb rule does not hold nonlinear responses – there is only linear term $a_1 = 1$. In neurophysiology situations [41] are known when a presynaptic synapse forms, i.e. when the axon of one neuron forms a synaptic bond with the axon of another. It corresponds to the retaining of only two first terms $r = 1, 2$ in the sum (1), given $a_r = 0$ at $r > 2$ in (2).

To simplify the expressions, let us not require the diagonal elements of the interconnection tensor to be strictly zero as in the Hopfield model. This requirement does not change the main results much: only some coefficients have to be redefined. In this case input signal $S_i$ can be written in the form

$$h_i = \sum_{\mu=1}^{M} \sum_{r=1}^{\infty} a_r x_i^{(\mu)} (\mathbf{x}_\mu \mathbf{y})^r \qquad (3)$$

where $\mathbf{y} = (y_1, y_2, ..., y_N)$ is the vector of the current network state.

To complete the picture, we should note that the energy of this sort of system has the form:

$$E = \sum_{\mu=1}^{M} \sum_{r=1}^{\infty} \frac{a_r}{r+1} (\mathbf{x}_\mu \mathbf{y})^{r+1} \qquad (4)$$

## PATTERN RECOGNITION

In the general case it is impossible to evaluate the pattern recognition efficiency. Let us consider the simplest case when the connections are determined by the following expression:

$$a_r = \frac{\alpha^r}{r!} \quad (5)$$

In this case expression (3) is reduced to the form:

$$h_i = \sum_m x_i^{(m)} \left( e^{\alpha \mathbf{x}_m \mathbf{y}} - 1 \right) \quad (6)$$

Considering (6) let us turn to the question of the recognition ability of the network under consideration. Let vector $\mathbf{y}$ arrive at the network input. The vector is a distorted version of pattern $\mathbf{x}_m$: $y_i = -x_i^{(m)}$ with probability $p$ and $y_i = x_i^{(m)}$ with probability $1-p$. Then (6) can be rewritten as

$$h_i = x_i^{(m)} \left( e^{\alpha \mathbf{x}_m \mathbf{y}} - 1 \right) + \sum_{\mu \neq m} x_i^{(\mu)} \left( e^{\alpha \mathbf{x}_\mu \mathbf{y}} - 1 \right) \quad (7)$$

where $\mathbf{x}_m \mathbf{y} = N(1-2p)$.

The condition for correct recognition is $h_i x_i^{(m)} > 0$ for all $i = 1,\ldots,N$ concurrently. In terms of spin systems this condition means that the direction of any spin $x_i^{(m)}$ is the same as that of acting local field $h_i$.

Let us consider two limiting cases of condition $h_i x_i^{(m)} > 0$.

a) In case $\alpha N \ll 1$ we come to the standard Hopfield model. Indeed, keeping in (7) only first-order terms in $\alpha$ we find that condition $h_i x_i^{(m)} > 0$ changes into the known relationship:

$$N(1-2p) + \sum_{\mu \neq m} \sum_{j \neq i} x_i^{(m)} x_i^{(\mu)} x_j^{(\mu)} y_j > 0 \quad (8)$$

The second term in the left part describes internal noise: normally distributed random variable with a zero mean and variance $\sigma_M^2 = NM$. Using probabilistic methods [28-30] we find that condition $h_i x_i^{(m)} > 0$ for one neuron is met with probability $Q = 1 - 0.5 \cdot \operatorname{erf} z$, where

$$z = \frac{N(1-2p)}{\sqrt{2}\sigma_M} = \sqrt{\frac{N}{2M}}(1-2p) \quad (9)$$

Hence the full probability of correct pattern recognition is $Q^N$, and the probability of incorrect recognition Pr has the form:

$$\Pr = 1 - \left( 1 - \frac{1}{2}\operatorname{erf} z \right)^N \quad (10)$$

Correct recognition corresponds to $z \to 0$. In this case the error probability can, to an accuracy of insignificant coefficients, be written as

$$\Pr \sim N \exp\left[\frac{N(1-2p)^2}{2M}\right] \tag{11}$$

From (11) and condition $\Pr \ll 1$ follows the well-known restriction on the associative memory capacity $M < M_{max}$, where

$$M_{max} = \frac{N}{2\ln N}(1-2p)^2 \tag{12}$$

Note that with $N \to \infty$ more correct expression $M_{max} \simeq 0.138N$ is obtained in [26], where the authors use methods of statistical physics to describe the Hopfield model with Hebb interconnection matrix. Paper [27] generalizes this result to the case of the weighed Hebb matrix.

b) At the limit $\alpha N \gg 1$ in (7) we can assume that the exponents are much greater than unit and ignore units. Then (7) takes the form:

$$h_i \approx x_i^{(m)} e^{\alpha \mathbf{x}_m \mathbf{y}} \left[1 + \sum_{\mu \neq m} x_i^{(m)} x_i^{(\mu)} e^{\alpha(\mathbf{x}_\mu - \mathbf{x}_m)\mathbf{y}}\right] \tag{13}$$

The sum in the square brackets plays the role of noise, i.e.

$$\text{Noise} = \sum_{\mu \neq m} x_i^{(m)} x_i^{(\mu)} e^{\alpha(\mathbf{x}_\mu - \mathbf{x}_m)\mathbf{y}} \tag{14}$$

Consequently, the condition of correct recognition $h_i x_i^{(m)} > 0$ is met when $\text{Noise} < 1$. It is easy to see that Noise is a normally distributed random variable with a zero mean. In the general case the variance of Noise is difficult to evaluate, yet it is not necessary (below we use a simpler approach).

For simplicity we consider the recognition of an undistorted pattern, i.e. $\mathbf{y} \equiv \mathbf{x}_m$ ($p = 0$). Let us denote the greatest measure of resemblance between different patterns as $\gamma N$: $\gamma N = \max(\mathbf{x}_\mu \mathbf{x}_\nu)$, $\mu, \nu = 1, 2, ..., M$. Then considering that $\mathbf{x}_m \mathbf{x}_\mu < \gamma N$ and $\mathbf{x}_m \mathbf{y} \equiv N$, we get from (14) that

$$\text{Noise} = \sum_{m \neq \mu} x_i^{(\mu)} x_i^{(m)} e^{\alpha \mathbf{x}_m \mathbf{x}_\mu - \alpha N} < \sum_{m \neq \mu} x_i^{(\mu)} x_i^{(m)} e^{-\alpha N(1-\gamma)} < \sum_{m \neq \mu} e^{-\alpha N(1-\gamma)} \tag{15}$$

Whence it follows that

$$\text{Noise} < M e^{-\alpha N(1-\gamma)} \tag{16}$$

If we assume that

$$\ln M < (1-\gamma)N\alpha \tag{17}$$

then for any $i \in \overline{1,N}$ noise will always be much less than signal ( Noise <1). Correspondingly, the memory capacity ( $M \leq M_{max}$ ) has the limit:

$$M_{max} = e^{\alpha N(1-\gamma)} \qquad (18)$$

As we see with $\alpha N \gg 1$ the associative memory capacity $M$ can be much bigger than the number of neurons $N$.

Expression (18) is found for the case when an undistorted pattern ( $p=0$ ) comes to the network input. By substituting $\mathbf{x}_m \mathbf{y} \equiv N(1-2p)$, we can generalize this expression to the case of recognition of a distorted pattern ( $p \neq 0$ ):

$$M_{max} = e^{\alpha N(1-2p-\gamma)} \qquad (19)$$

As we see, condition $1-2p > \gamma$ is enough to achieve effective recognition. The condition means that distortions are not too big: the opposite condition $\gamma > 1-2p$ implies that heavy distortions make pattern $\mathbf{x}_m$ be like one of stored patterns $\mathbf{x}_\mu$ ( $\mu \neq m$ ). Of course, in this case pattern $\mathbf{x}_\mu$ rather than pattern $\mathbf{x}_m$ will be recognized.

## DISCUSSION

The above considerations demonstrate that nonlinear interconnections with coefficients like (5) lead to a considerable increase in network recognition efficiency: the memory capacity and basin of attraction grow significantly with nonlinearity parameter $\alpha$. As follows from (19), when $\alpha N \to \infty$, the network can reliably recognize patterns with distortions $p \to 0.5$. At the same time the associative memory capacity, according to (19), grows exponentially with $N$. And this is not hardly surprising because both the network capacity and the network recognition ability are determined by the number of connections rather than the number of neurons: in the limit case $\alpha N \to \infty$ the number of connections also approaches infinity.

The model determined by expressions (1) – (4) can hardly offer a correct description of hardware realization of memristor crossbar-based neural networks. However, it offers an idea how parasitic currents can influence network recognition efficiency. It is most likely that considering only two terms in sum (1) is enough to describe this sort of influence, which means that (1) should be written in the form:

$$h_i = \sum_{j \neq i} T_{ij}^{(1)} y_i y_j + \sum_{j,k \neq i}^{\infty} T_{ijk}^{(2)} y_i y_j y_k \qquad (20)$$

where the first term describes the standard Hopfield model, and the second term determines the cubic nonlinearity (the presence of presynaptic synapses) caused by noise signals:

$$T_{ij}^{(1)} = \sum_{\mu=1}^{M} x_i^{(\mu)} x_j^{(\mu)}, \quad T_{ijk}^{(2)} = a \sum_{\mu=1}^{M} x_i^{(\mu)} x_j^{(\mu)} x_k^{(\mu)} \qquad (21)$$

To check the effect of the second term on recognition efficiency, the function of a small network ($N=100$) was modelled. For each parameter $a$, distortion parameter $p$ and load parameter $M/N$ we generated $10^3$ Hebbian matrices and calculated average $\text{Pr}$. The results of the modelling are shown in Figures 1 and 2.

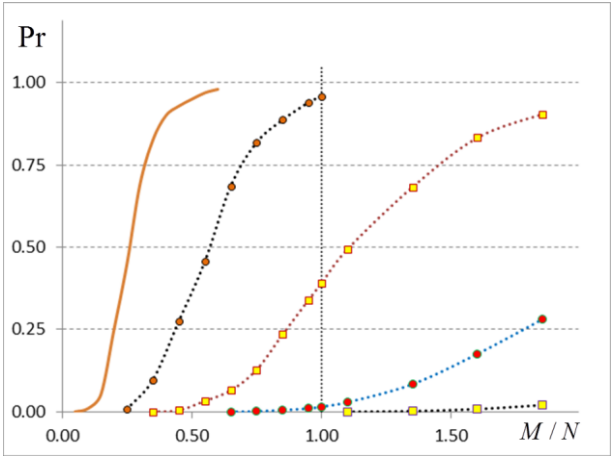
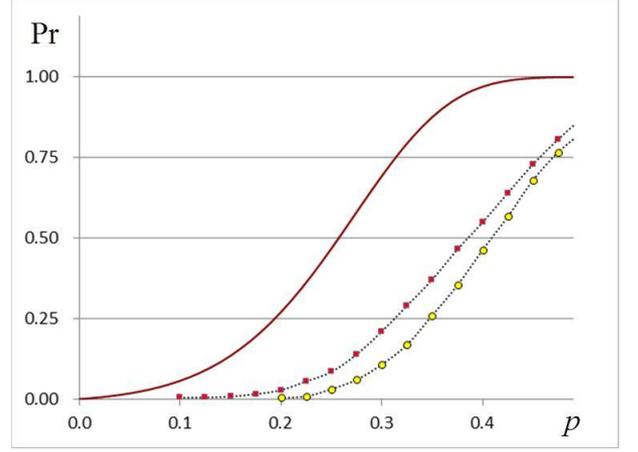

Fig. 1. Relations between recognition error $\text{Pr}$ and load parameter $M/N$. Curves from left to right correspond to $a = 0, 0.01, 0.02, 0.05, 0.1$.

Fig. 2. Relations between recognition error $\text{Pr}$ and distortion parameter $p$. Curves from left to right correspond to $a = 0, 0.01, 0.1$. The load parameter is fixed: $M = 0.13N$, $N = 100$.

Figure 1 shows the relationship between recognition error $\text{Pr}$ and load parameter $M/N$ in case of undistorted patterns ($p=0$). As we see, the standard Hopfield network (solid-line curve, $a=0$) stops recognizing an undistorted pattern even at low loads ($M/N \geq 0.07$). At the same time the introduction of even small nonlinearity ($a = 0.01$) guarantees reliable pattern recognition at $M/N \leq 0.25$. As parameter $a$ grows, the memory capacity increases rapidly, e.g. at $a = 0.1$ a non-zero error appears only at $M > 1.2N$.

Figure 2 shows the relationship between recognition error $\text{Pr}$ and pattern distortion parameter $p$ at the given load parameter $M/N = 0.13$. It is seen that the standard Hopfield network (solid-line curve, $a=0$) has a small basin of attraction, errors start occurring even at very low distortions $p \sim 0.05$. At the same time, the addition of small nonlinearity significantly increases the basin of attraction: when $a = 0.01$ the network begins to fail at $p \geq 0.1$; when $a = 0.1$, a non-zero error appears only at $p \geq 0.2$.

FUNDING

The work is supported by the State Program of Research Center *Kurchatov Institute* – SRISA, project no. FNEF-2024-0001.

REFERENCES

[1] Pershin Y. and Di Ventra M. Experimental demonstration of associative memory with memristive neural networks. Neural Networks, 2010, vol. 23, no. 7, pp. 881–886.

[2] Jo S. H., Chang T., Ebong I., et al. Nanoscale memristor device as synapse in neuromorphic systems. Nanoletters, 2010, vol. 10, no. 4, pp. 1297–1301.

[3] Truong S. N., Ham S.-J., Min K.-S. Neuromorphic crossbar circuit with nanoscale filamentary-switching binary memristors for speech recognition. Nanoscale Res. Lett., 2014, no. 9:629.

[4] Zhu X., Yang X., Wu C., et al. Hamming network circuits based on CMOS/memristor hybrid design. IEICE Electronics Express, 2013, vol. 10, no. 12, pp. 1–9.

[5] A. N. Palagushkin; D. V. Roshchupkin; F. A. Yudkin;D. V. et al. Aspects of the a-TiOx memristor active medium technology. J. Appl. Phys. 124, 205109 (2018)

[6] Kim H., Mahmoodi M.R., Nili H., Strukov D. B. 4K-memristor analog-grade passive crossbar circuit. Nat Commun 12, 5198 (2021). https://doi.org/10.1038/s41467-021-25455-0.

[7] X. Liu, Z. Zeng. Memristor crossbar architectures for implementing deep neural networks. Complex & Intelligent Systems (2022) 8:787–802.

[8] Kaneko, Y., Nishitani, Y. & Ueda, M. Ferroelectric artificial synapses for recognition of a multishaded image. IEEE Trans. Electron Devices 61, 2827–2833 (2014).

[9] Boyn, S. et al. Learning through ferroelectric domain dynamics in solid-state synapses. Nat. Commun. 8, 14736 (2017).

[10] Romera, M. et al. Vowel recognition with four coupled spin-torque nanooscillators. Nature 563, 230–234 (2018).

[11] Fuller, E. J. et al. Parallel programming of an ionic floating-gate memory array for scalable neuromorphic computing. Science 364, 570–574 (2019).

[12] Goswami, S. et al. Robust resistive memory devices using solution-processable metal-coordinated azo aromatics. Nat. Mater. 16, 1216–1224 (2017).

[13] Indiveri, G. et al. Integration of nanoscale memristor synapses in neuromorphic computing architectures. Nanotechnology 24, 384010 (2013).

[14] Prezioso, M. et al. Training and operation of an integrated neuromorphic network based on metal-oxide memristors. Nature 521, 61–64 (2015).

[15] Ambrogio, S. et al. Neuromorphic learning and recognition with onetransistor-one-resistor synapses and bistable metal oxide RRAM. IEEE Trans. Electron Devices 63, 1508–1515 (2016).

[16] Adam, G. C. et al. 3-D memristor crossbars for analog and neuromorphic computing applications. IEEE Trans. Electron Devices 64, 312–318 (2017).

[17] Hu, M. et al. Memristor-based analog computation and neural network classification with a dot product engine. Adv. Mater. 30, 1705914 (2018).

[18] Merrikh Bayat, F. et al. Implementation of multilayer perceptron network with highly uniform passive memristive crossbar circuits. Nat. Commun. 9, 2331 (2018).

[19] Chakrabartty, S. & Cauwenberghs, G. Sub-microwatt analog VLSI trainable pattern classifier. IEEE J. Solid-State Circuits 42, 1169–1179 (2007).

[20] Ramakrishnan, S. & Hasler, J. Vector-matrix multiply and winner-take-all as an analog classifier. IEEE Trans. Very Large Scale Integr. Syst. 22, 353–361 (2014).

[21] Merrikh Bayat, F. et al. High-performance mixed-signal neurocomputing with nanoscale floating-gate memory cells. IEEE Trans. Neural Netw. Learn. Syst. 29, 4782–4790 (2018).

[22] Eryilmaz, S. B. et al. Brain-like associative learning using a nanoscale nonvolatile phase change synaptic device array. Front. Neurosci. 8, 205 (2014).

[23] Burr, G. W. et al. Experimental demonstration and tolerancing of a large-scale neural network (165 000 synapses) using phase-change memory as the synaptic weight element. IEEE Trans. Electron Devices 62, 3498–3507 (2015).

[24] Boybat, I. et al. Neuromorphic computing with multi-memristive synapses. Nat. Commun. 9, 2514 (2018).

[25] Hopfield, J.J. Neural Networks and physical systems with emergent collective computational abilities. Proc. Nat. Acad. Sci.USA. v.79, pp.2554-2558 (1982).

[26] Amit, D. J.; Gutfreund, H.; and Sompolinsky, H. Storing infinite numbers of patterns in a spin-glass model of neural networks. Physical Review Letters, 55(14): 1530–1533 (1985).

[27] Karandashev, I.; Kryzhanovsky B.; Litinskii L. Weighted patterns as a tool to improve the Hopfield model. Physical Review E 85, 041925 (2012) .

[28] Kryzhanovsky, B.V.; Kryzhanovsky, V.M.; Mikaelian, A.L. Parametric dynamic neural network recognition power. Optical Memory&Neural Network, Vol. 10, №4, pp.211-218 (2001).

[29] Kryzhanovsky, B.V.; Mikaelian, A.L. Doklady Mathematics, vol.65, No.2, pp. 286-288 (2002). On the Recognition Ability of a Neural Network on Neurons with Parametric Transformation of Frequencies.

[30] Kryzhanovsky, B.; Litinskii, L.; Fonarev, A. An Effective Associative Memory for Pattern Recognition. Lecture Notes in Computer Science, Vol.2810/2003, pp.179-186 (2003).

[31] Kryzhanovsky, B.V.; Kryzhanovsky, V.M. Modified q-state Potts Model with Binarized Synaptic Coefficients. Lecture Notes in Computer Science, Springer Berlin / Heidelberg , Vol. 5164/2008, pp.72-80.

[32] Karandashev, I.; Kryzhanovsky B.; Litinskii L. Properties of the Hopfield Model with Weighted Patterns // Lecture Notes in Computer Science, vol. 7552, Part I, pp. 41-48 (2012). Springer Berlin/Heidelberg.

[33] Agliari, E.; Albanese, L.; Alemanno, F.; et al. 2023a. Dense Hebbian neural networks: A replica symmetric picture of supervised learning. Physica A: Statistical Mechanics and its Applications, 626: 129076.

[34] Barra, A.; Bernacchia, A.; Santucci, E.; and Contucci, P. 2012. On the equivalence of hopfield networks and boltzmann machines. Neural Networks, 34: 1–9.

[35] Bertram, R.; and Rubin, J. E. 2017. Multi-timescale systems and fast-slow analysis. Mathematical biosciences, 287: 105–121.

[36] Fanaskov, V. and Oseledets, I. 2024. Associative memory and dead neurons. arXiv:2410.13866.

[37] Hinton, G. E. 2012. A practical guide to training restricted boltzmann machines. Lecture Notes in Science, 7700 LECTU: 599–619.

[38] Hoover, B.; Chau, D. H.; Strobelt, H.; et al. 2022. A Universal Abstraction for Hierarchical Hopfield Networks. In The Symbiosis of Deep Learning and Differential Equations II.

[39] Krotov, D., Hopfield, J. 2018. Dense associative memory is robust to adversarial inputs. Neural Computation, 30(12): 3151–3167.

[40] Hebb, D.O. The Organization of Behavior. A Neuropsychological Theory, NY: Wiley, 1949.

[41] Eccles, J. C. The physiology of nerve cells. Baltimore, The Hopkins press, 1957.